\documentclass[final,5p,times,twocolumn]{elsarticle}

\newif\ifcountpages
\countpagesfalse

\usepackage{graphicx}
\usepackage{caption}
\usepackage{amssymb}
\usepackage{chngpage}
\usepackage{atlasphysics}
\RequirePackage[colorlinks,breaklinks,pdftitle={ATLAS draft},pdfauthor={The ATLAS Collaboration}]{hyperref}  
\hypersetup{linkcolor=blue,citecolor=blue,filecolor=black,urlcolor=blue} 
\usepackage{hypernat}

\newcommand\lumiseventev{4.5}
\newcommand\lumieighttev{20.3}

\newcommand\pvaluelow{0.05}

\newcommand\pvaluehiggs{0.27}

\newcommand\sigmahigh{1.6}
\newcommand\sigmahiggs{0.6}
\newcommand\pvaluelowexp{0.34}
\newcommand\pvaluehighexp{0.44}
\newcommand\pvaluehiggsexp{0.42}
\newcommand\sigmalowexp{0.2}

\newcommand\sigmahiggsexp{0.2}
\newcommand\excllowexp{5}
\newcommand\exclhighexp{15}
\newcommand\exclhiggsexp{9}
\newcommand\excllow{3.5}
\newcommand\exclhigh{18}
\newcommand\exclhiggs{11} 

\newcommand{\Etiso}   {\ensuremath{E_{\mathrm{T}}^{\mathrm{iso}}}}

\biboptions{numbers,square,comma,sort&compress}
\journal{Physics Letters B}

\usepackage{preprintcover}  
\PreprintCoverPaperTitle{Search for Higgs boson decays to a photon and a $Z$ boson in $pp$ collisions at $\sqrt{s}=7$ and 8 TeV with the ATLAS detector}  
\PreprintIdNumber{CERN-PH-EP-2014-006}  
\PreprintCoverAbstract{A search is reported for a neutral Higgs boson in the decay channel
$H\to Z\gamma$, $Z\to\ell^+\ell^-$ ($\ell=e,\mu$),
using \lumiseventev~fb$^{-1}$ of $pp$ collisions at $\sqrt{s} = 7$~TeV and
\lumieighttev~fb$^{-1}$ of $pp$ collisions at $\sqrt{s} = 8$~TeV, recorded
by the ATLAS detector at the CERN Large Hadron Collider.
The observed distribution of the invariant mass of the three
final-state particles, $m_{\ell\ell\gamma}$, is consistent with the
Standard Model hypothesis in the investigated mass range of 120--150 GeV.
For a Higgs boson with a mass of 125.5 GeV,
the observed upper limit at the 95\% confidence level is \exclhiggs\
times the Standard Model expectation.
Upper limits are set on the cross section times branching ratio of a neutral
Higgs boson with mass in the range 120--150 GeV
between 0.13 and 0.5 pb for $\sqrt{s}=8$ TeV at 95\%
confidence level.
}  
\PreprintJournalName{Physics Letters B}  

\clearpage

\begin{document}

\begin{frontmatter}

\title{Search for Higgs boson decays to a photon and a $Z$ boson in $pp$ collisions at $\sqrt{s}=7$ and 8 TeV with the ATLAS detector}

\author{ATLAS Collaboration, G. Aad \textit{et al.} (full author list given at the end of the article in Appendix)\\
}

\begin{abstract}

A search is reported for a neutral Higgs boson in the decay channel 
$H\to Z\gamma$, $Z\to\ell^+\ell^-$ ($\ell=e,\mu$),
using \lumiseventev~fb$^{-1}$ of $pp$ collisions at $\sqrt{s} = 7$~TeV and 
\lumieighttev~fb$^{-1}$ of $pp$ collisions at $\sqrt{s} = 8$~TeV, recorded
by the ATLAS detector at the CERN Large Hadron Collider.
The observed distribution of the invariant mass of the three
final-state particles, $m_{\ell\ell\gamma}$, is consistent with the
Standard Model hypothesis in the investigated mass range of 120--150 GeV.
For a Higgs boson with a mass of 125.5 GeV, 
the observed upper limit at the 95\% confidence level is \exclhiggs\
times the Standard Model expectation.
Upper limits are set on the cross section times branching ratio of a neutral
Higgs boson with mass in the range 120--150 GeV 
between 0.13 and 0.5 pb for $\sqrt{s}=8$ TeV at 95\%
confidence level.

\end{abstract}


\end{frontmatter}

\section{Introduction}
\label{sec:intro}
In July 2012 a new particle decaying to dibosons ($\gamma\gamma$, $ZZ$, $WW$)
was discovered by the ATLAS~\cite{ATLAS_Higgs} and 
CMS~\cite{CMS_Higgs} experiments at the CERN Large Hadron Collider (LHC).
The observed properties of this particle, such as its couplings 
to fermions and bosons~\cite{ATLAS_couplings,CMS_couplings} and its spin and
parity~\cite{ATLAS_spin,CMS_spin}, are consistent with those of a
Standard Model (SM) Higgs boson with a mass 
near 125.5 GeV~\cite{ATLAS_couplings}.

This Letter presents a search for a Higgs boson $H$ decaying
to $Z\gamma$, $Z\to\ell^+\ell^-$ ($\ell=e,\mu$),\footnote{In the following $\ell$ denotes
either an electron or a muon, and the charge of the leptons
is omitted for simplicity.} using
$pp$ collisions at $\sqrt{s}=7$ and 8~TeV recorded
with the ATLAS detector at the LHC during 2011 and 2012. 
The Higgs boson is assumed to have SM-like spin and
production properties and a mass between 120 and 150 GeV.
The integrated luminosity presently available 
enables the exclusion of large anomalous couplings
to $Z\gamma$, compared with the SM prediction.
The signal is expected to yield a narrow peak in the 
reconstructed $\ell\ell\gamma$ invariant-mass distribution 
over a smooth background dominated by continuum 
$Z{+}\gamma$ production, $Z\to \ell\ell\gamma$ radiative
decays and $Z+$jets events where a jet is misidentified as a photon.
A similar search was recently published by the CMS
Collaboration~\cite{CMS-HZG}, which set an upper limit of 9.5
times the SM expectation, at 95\% confidence level (CL),
on the $pp\to H\to Z\gamma$ cross section for $m_H=125$ GeV.

In the SM, the Higgs boson is produced mainly through five production 
processes: gluon fusion (ggF), vector-boson fusion (VBF), and
associated production with either a $W$ boson ($WH$), a $Z$ boson $(ZH)$ or 
a $t\bar{t}$ pair ($t\bar{t}H$)~\cite{LHCHiggsCrossSectionWorkingGroup:2011ti,LHCHiggsCrossSectionWorkingGroup:2012vm,Heinemeyer:2013tqa}. 
For a mass of 125.5 GeV the SM $pp\rightarrow H$ cross section 
is $\sigma=22$ (17) pb at $\sqrt{s}=8$ (7) TeV.
Higgs boson decays to $Z\gamma$ in the SM proceed through loop diagrams
mostly mediated by $W$ bosons, similar to $H\rightarrow\gamma\gamma$.
The $H\to Z\gamma$ branching ratio of a SM Higgs boson with a mass of 125.5 GeV 
is $B(H\to Z\gamma)=1.6\times 10^{-3}$ compared to $B(H\rightarrow\gamma\gamma)=2.3\times 10^{-3}$.
The branching fractions of the Z to leptons leads
to a $pp\to H\rightarrow \ell\ell\gamma$ 
cross section of 2.3 (1.8) fb at 8 (7) TeV, similar to that of
$pp\rightarrow H\rightarrow ZZ^*\to 4\ell$ and only 5\% of that of
$pp\rightarrow H\rightarrow \gamma\gamma$. 

Modifications of the $H\to Z\gamma$ coupling with respect to the SM
prediction are expected if $H$ is a neutral scalar 
of a different origin~\cite{low,low2} 
or a composite state~\cite{Azatov:2013ura},
as well as in models with additional colourless charged scalars,
leptons or vector bosons coupled to the Higgs boson and 
exchanged in the $H\to Z\gamma$ loop~\cite{extensions, Chen:2013vi, carena}.
A determination of both the $H\to \gamma\gamma$ and 
$H\to Z\gamma$ decay rates can 
help to determine whether the newly discovered Higgs boson is 
indeed the one predicted in the SM, or provide information 
on the quantum numbers of the new particles
exchanged in the loops or on the compositeness scale.
While constraints from the observed rates in the other final states,
particularly the diphoton channel, typically limit the expected
$H\to Z\gamma$ decay rate in the models mentioned above to be within a
factor of two of the SM expectation, larger enhancements can be
obtained in some scenarios by careful parameter choices~\cite{Azatov:2013ura,extensions}. 

\section{Experimental setup and dataset}
\label{sec:AtlasDet}

The ATLAS detector~\cite{Aad:2008zzm} is a multi-purpose particle
detector with approximately forward-backward symmetric
cylindrical geometry.\footnote{ATLAS uses a right-handed coordinate
  system with its origin at the nominal interaction point (IP) in the
  centre of the detector and the $z$-axis along the beam pipe. The
  $x$-axis points from the IP to the centre of the LHC ring, and the
  $y$-axis points upward. Cylindrical coordinates $(r,\phi)$ are used
  in the transverse plane, $\phi$ being the azimuthal angle around the
  beam pipe. The pseudorapidity is defined in terms of the polar angle
  $\theta$ as $\eta=-\ln\tan(\theta/2)$.} 
The inner tracking detector (ID) covers $\left| \eta \right|<2.5$ and
consists of a silicon pixel detector, a silicon microstrip detector, and a
transition radiation tracker. 
The ID is surrounded by a thin
superconducting solenoid providing a $2\:\rm{T}$ axial magnetic field
and by a high-granularity lead/liquid-argon (LAr) sampling
electromagnetic calorimeter. The electromagnetic
calorimeter measures the energy and the position of
electromagnetic showers with $\left| \eta \right|<3.2$. 
It includes a presampler (for $|\eta|<1.8$) and three sampling layers, 
longitudinal in shower depth, up to $|\eta|<2.5$.
LAr sampling calorimeters are also used to measure hadronic showers in
the end-cap ($1.5<\left| \eta \right|<3.2$) and forward ($3.1<\left|
\eta \right|<4.9$) regions, while an iron/scintillator tile
calorimeter measures hadronic showers in the central
region ($\left| \eta \right|<1.7$).  The muon spectrometer (MS)
surrounds the calorimeters and consists of three
large superconducting air-core toroid magnets, each with eight coils,
a system of precision tracking chambers ($\left| \eta \right|<2.7$),
and fast tracking chambers ($\left| \eta \right|<2.4$) for triggering.  
A three-level trigger system selects events to be recorded 
for offline analysis. 

Events are collected using the lowest threshold unprescaled single-lepton
or dilepton triggers~\cite{Trigger}. 
For the single-muon trigger the transverse momentum, $\pT$, threshold
is 24 (18) GeV for $\sqrt{s}=8$ (7) TeV, while for the single-electron
trigger the transverse energy, $\ET$, threshold is 25 (20) GeV. 
For the dimuon triggers the thresholds are $\pT> 13$ (10)~GeV for
each muon,  while for the dielectron triggers the thresholds are 
$\ET> 12$ GeV for each electron. 
At $\sqrt{s}=8$ TeV a dimuon trigger is also used with asymmetric 
thresholds $p_{\rm T1} > 18$ GeV and $p_{\rm T2} > 8$ GeV.
The trigger efficiency with respect to events satisfying the selection 
criteria is 99\% in the $ee\gamma$ channel and 92\% in the
$\mu\mu\gamma$ channel due to the reduced geometric acceptance of the muon
trigger system in the $|\eta|<1.05$ and $|\eta|>2.4$ region.
Events with data quality problems are discarded. 
The integrated luminosity after the trigger and data quality
requirements corresponds to \lumieighttev~\ifb\ (\lumiseventev~\ifb)~\cite{lumi2011}
at $\sqrt{s} = 8$ (7) TeV.

\section{Simulated samples}
\label{sec:SimSamples}
The event generators used to model SM signal and background processes 
in samples of Monte Carlo (MC) simulated events are listed in
Table~\ref{tab:generators}.

\begin{table}[!htbp] 
  \begin{center}
    \caption{Event generators used to model the signal (first two
      rows) and background (last four rows) processes.} 
    \label{tab:generators}
    \begin{tabular}{cc}
      \hline\hline
      Process & Generator \\
      \hline
      ggF, VBF                & POWHEG~\cite{powheg_ggF,powheg_VBF,Bagnaschi:2011tu}+PYTHIA8~\cite{Pythia8} \\
      $WH$, $ZH$, $t\bar{t}H$ & PYTHIA8 \\
      \hline
      $Z{+}\gamma$ and $Z\to\ell\ell\gamma$ & SHERPA~\cite{sherpa2,sherpa3} \\
      $Z+$jets                           & SHERPA, ALPGEN~\cite{Mangano:2002ea}+HERWIG~\cite{Herwig} \\
      $t\bar{t}$                         & MC@NLO~\cite{Frixione:2002ik,Frixione:2003ei}+HERWIG \\
      $WZ$                               & SHERPA, POWHEG+PYTHIA8 \\
      \hline\hline
    \end{tabular}
  \end{center}
\end{table}

The $H\rightarrow Z\gamma$ signal from the dominant ggF and
VBF processes, corresponding to 95\% of the SM production cross section, is generated with
POWHEG, interfaced to PYTHIA 8.170 for showering and hadronisation, using the
CT10 parton distribution functions (PDFs)~\cite{Lai:2010vv}.
Gluon-fusion events are reweighted to match the Higgs boson $\pT$ 
distribution predicted by HRES2~\cite{HRES2}.
The signal from associated production ($WH$, $ZH$ or $t\bar{t}H$) 
is generated with PYTHIA 8.170 using the CTEQ6L1 PDFs~\cite{JHEP_0207_012}.
Signal events are generated for Higgs boson masses $m_H$ between 120 and
150 GeV, in intervals of 5 GeV, at both $\sqrt{s}=7$~TeV and 
$\sqrt{s}=8\:\tev$. 
For the same value of the mass, events corresponding to 
different Higgs boson production modes are combined according to their respective
SM cross sections.
\\
\indent
The predicted SM cross sections and
branching ratios are compiled in
Refs.~\cite{LHCHiggsCrossSectionWorkingGroup:2011ti,
  LHCHiggsCrossSectionWorkingGroup:2012vm,Heinemeyer:2013tqa}.
The production cross sections are computed at 
next-to-next-to-leading order in the strong coupling constant
$\alpha_s$ and at next-to-leading order (NLO) in the electroweak
coupling constant $\alpha$, except for 
the $t\bar{t}H$ cross section, which is calculated at
NLO in $\alpha_s$~\cite{Harlander:2002wh,
  Anastasiou:2002yz, Ravindran:2003um, Anastasiou:2012hx,
  deFlorian:2012yg,Aglietti:2004nj,Actis:2008ug,
  Brein:2003wg,Ciccolini:2003jy,Beenakker:2002nc, Dawson:2003zu}.
Theoretical uncertainties on the production cross section arise
from the choice of renormalisation and factorisation scales in the
fixed-order calculations as well as the uncertainties on the PDFs and the value
of $\alpha_s$ used in the perturbative expansion.
They depend only mildly on the centre-of-mass energy and on the Higgs
boson mass in the range $120<m_H<150$~GeV.
The scale uncertainties are uncorrelated among the five Higgs boson
production modes that are considered; 
for $m_H=125.5$~GeV at $\sqrt{s}=8$ TeV, they amount to
$^{+7}_{-8}\%$ for ggF, $\pm 0.2\%$ for VBF, $\pm 1\%$ for $WH$,
$\pm 3\%$ for $ZH$ and $^{+4}_{-9}\%$ for $t\bar{t}H$.
PDF+$\alpha_s$ uncertainties are correlated among the gluon-fusion and
$t\bar tH$ processes, which are initiated by gluons,
and among the VBF and $WH/ZH$ processes, which are initiated by quarks;
for $m_H=125.5$~GeV at $\sqrt{s}=8$ TeV, the uncertainties are 
around $\pm 8$\% for $gg\to H$ and $t\bar tH$ and 
around $\pm 2.5$\% for the other three Higgs boson production modes.
The Higgs boson branching ratios are computed using the HDECAY and 
Prophecy4f programs~\cite{Djouadi:1997yw,Bredenstein:2006rh,
  Actis:2008ts}. 
The relative uncertainty on the $H\to Z\gamma$ branching ratio
varies between $\pm 9$\% for $m_H=120$ GeV and $\pm 6$\% for
$m_H=150$ GeV. An additional $\pm5\%$~\cite{Dicus:2013lta} 
accounts for the effect, in the 
selected phase space of the $\ell\ell\gamma$ final state, of the
interfering $H\to\ell\ell\gamma$ decay amplitudes 
that are neglected in the calculation of
Refs.~\cite{LHCHiggsCrossSectionWorkingGroup:2011ti, 
  LHCHiggsCrossSectionWorkingGroup:2012vm,Heinemeyer:2013tqa}.
They originate from internal
photon conversion in Higgs boson decays to diphotons
($H\to\gamma^*\gamma\to\ell\ell\gamma$) or from radiative
Higgs boson decays to dileptons 
($H\to\ell\ell^*\to\ell\ell\gamma$ in the $Z$ mass window)~\cite{Chen:2012ju,Firan:2007tp}.
\newline
\indent
Various background samples are also generated: they are used
to study the background parameterisation and possible systematic 
biases in the fit described in Section~\ref{sec:Results}
and not to extract the final result.
The samples produced with ALPGEN or MC@NLO are interfaced to HERWIG
6.510~\cite{Herwig} for parton showering, fragmentation into 
particles and to model the underlying event, using JIMMY
4.31~\cite{jimmy} to generate multiple-parton interactions. 
The SHERPA, MC@NLO and POWHEG samples are generated using the CT10 PDFs, while 
the ALPGEN samples use the CTEQ6L1 ones.
\newline
\indent All Monte Carlo samples are processed through a complete
simulation of the ATLAS detector response~\cite{ATLASsimulation}
using \textsc{Geant4}~\cite{geant}.
Additional $pp$ interactions in the same and nearby bunch crossings
(pile-up) are included in the simulation. 
The MC samples are reweighted to reproduce the distribution
of the mean number of interactions per bunch crossing 
(9 and 21 on average in the data taken at $\sqrt{s}=7$ 
and 8 TeV, respectively)
and the length of the luminous region observed in data.

\section{Event selection and backgrounds}
\label{sec:EvSel}

\subsection{Event selection}
\label{ss:EvSel}
Events are required to contain at least one primary vertex, 
determined from a fit to the tracks reconstructed
in the inner detector and consistent with a common origin.
The primary vertex
with the largest sum of the squared transverse momenta 
of the tracks associated with it is considered as the
primary vertex of the hard interaction.

The selection of leptons and photons is similar to that
used for the $H\to\gamma\gamma$ and $H\to 4\ell$ 
measurements~\cite{ATLAS_Higgs}, the main difference
being the minimum transverse momentum threshold.
Events are required to contain at least one photon and two opposite-sign
same-flavour leptons.

Muon candidates are formed from tracks reconstructed either in the ID
or in the MS~\cite{muonreco}.
They are required to have transverse momentum 
$\pt>10$~GeV and $|\eta| < 2.7$. 
In the central barrel region $|\eta| < 0.1$, 
which lacks MS coverage, ID tracks 
are identified as muons based on the
associated energy deposits in the calorimeter. 
These candidates must have $\pt>15$~GeV.
The inner detector tracks associated with muons that are 
identified inside the ID acceptance are required to have
a minimum number of associated hits in each of the ID sub-detectors
(to ensure good track reconstruction) and to have 
transverse (longitudinal) impact parameter $d_0$ ($z_0$), with respect
to the primary vertex, smaller than 1~mm (10 mm).  

Electrons and photons are reconstructed from clusters 
of energy deposits in the electromagnetic
calorimeter~\cite{PhotonPerformanceNote}.
Tracks matched to electron candidates (and, for 8 TeV data, 
from photon conversions) and having enough associated hits in the
silicon detectors are fitted using a Gaussian-Sum Filter, which 
accounts for bremsstrahlung energy loss~\cite{GSFConf}.

Electron candidates are required to have a transverse energy greater
than 10 GeV, pseudorapidity $|\eta|<2.47$, and a well-reconstructed ID
track pointing to the electromagnetic calorimeter cluster.
The cluster should satisfy a set of identification criteria that
require the longitudinal and transverse shower profiles to be consistent
with those expected for electromagnetic showers~\cite{Aad:2011mk}. 
The electron track is required to have a hit in the innermost
pixel layer of the ID when passing through an active module and is also
required to  have a longitudinal impact parameter, with respect to the
primary vertex, smaller than 10 mm. 

Photon candidates are required to have a transverse energy
greater than 15 GeV and pseudorapidity within the regions $|\eta|<1.37$ or
$1.52<|\eta|<2.37$, where the first calorimeter layer has high
granularity.
Photons reconstructed in or near regions of the calorimeter affected by 
read-out or high-voltage failures are not accepted.
The identification of photons is performed through a cut-based
selection based on shower shapes measured in the first two
longitudinal layers of the electromagnetic calorimeter and on the
leakage into the hadronic calorimeter~\cite{Aad:2010sp}.
To further suppress hadronic background, 
the calorimeter isolation transverse energy \Etiso~\cite{ATLAS_Higgs}
in a cone of size $\Delta R = \sqrt{(\Delta\eta)^2+(\Delta\phi)^2}=0.4$ 
around the photon candidate is required to be lower than $4\gev$,
after subtracting the contributions from the photon itself and 
from the underlying event and pile-up.

Removal of overlapping electrons and muons that satisfy all selection
criteria and share the same inner detector track 
is performed: if the muon is identified by the 
MS, then the electron candidate is discarded; otherwise the muon
candidate is rejected.
Photon candidates within a $\Delta R = 0.3$ cone of a
selected electron or muon candidate are also rejected, thus 
suppressing background from $Z\to\ell\ell\gamma$ events and 
signal from radiative Higgs boson decays to dileptons.

$Z$ boson candidates are reconstructed from pairs of same-flavour,
opposite-sign leptons passing the previous selections. 
At least one of the two muons from $Z\to\mu\mu$ must be
reconstructed both in the ID and the MS.

Higgs boson candidates are reconstructed from the combination of a
$Z$ boson and a photon candidate. In each event only the $Z$ candidate 
with invariant mass closest to the $Z$ pole mass and the
photon with largest transverse energy are retained.
In the selected events, the triggering leptons are required 
to match one (or in the case of dilepton-triggered events, both) 
of the $Z$ candidate's leptons.
Track and calorimeter isolation requirements,
as well as additional track impact parameter selections,
are also applied to the leptons forming the $Z$ boson 
candidate~\cite{ATLAS_Higgs}.
The track isolation $\sum \pT$, inside a $\Delta R = 0.2$ cone
around the lepton, excluding the lepton track, divided by the lepton \pT,
must be smaller than 0.15.
The calorimeter isolation for electrons, computed similarly to
\Etiso~for photons but with $\Delta R = 0.2$,
divided by the electron \ET, must be lower than
0.2.
Muons are required to have a normalised calorimeter
isolation $\ET^{\rm cone}/\pT$ less than 0.3 (0.15 in the case of muons 
without an ID track) inside a $\Delta R = 0.2$ cone
around the muon direction.
For both the track- and calorimeter-based isolation any contributions
due to the other lepton from the candidate $Z$ decay are
subtracted.
The transverse impact parameter significance $|d_0|/\sigma_{d_0}$ of the
ID track associated with a lepton within the acceptance of the inner
detector is required to be less than 3.5 and 6.5 for muons and electrons,
respectively.
The electron impact parameter is affected by bremsstrahlung 
and it thus has a broader distribution.

Finally, the dilepton invariant mass ($m_{\ell\ell}$) 
and the invariant mass of the 
$\ell\ell\gamma$ final-state particles ($m_{\ell\ell\gamma}$) are required to
satisfy $m_{\ell\ell}>m_Z-10$~GeV and $115<m_{\ell\ell\gamma}<170$~GeV, respectively.
These criteria further suppress events from 
$Z\rightarrow \ell\ell\gamma$, 
as well as reducing the contribution to the signal from 
internal photon conversions in $H\to\gamma\gamma$
and radiation from leptons in $H\to\ell\ell$ to a negligible level~\cite{Dicus:2013lta}.
The number of events satisfying all the selection criteria in
 $\sqrt{s}=8\:\tev$ ($\sqrt{s}=7\:\tev$) data is
7798 (1041) in the $Z\rightarrow ee$ channel and
9530 (1400) in the $Z\rightarrow \mu\mu$ channel.

The same reconstruction algorithms and selection criteria are
used for simulated events.
The simulation is corrected to take into account measured data-MC
differences in photon and lepton efficiencies
and energy or momentum resolution.
The acceptance of the kinematic requirements for simulated 
$H\to Z\gamma\to\ell\ell\gamma$ signal events at $m_H=125.5$ GeV
is 54\% for $\ell=e$ and 57\% for $\ell=\mu$, due to the larger 
acceptance in muon pseudorapidity.
The average photon reconstruction and selection efficiency is 
68\% (61\%) while the $Z\to \ell\ell$ reconstruction and selection 
efficiency is 74\% (67\%) and 88\% (88\%)
for $\ell=e$ and $\ell=\mu$, respectively, at $\sqrt{s}=8$ (7) TeV.
The larger photon and electron efficiencies in 8 TeV data are due to a
re-optimisation of the photon and electron identification criteria 
prior to the 8 TeV data taking.
Including the acceptance and the reconstruction, selection and trigger
efficiencies, the overall signal efficiency for 
$H\to Z\gamma\to\ell\ell\gamma$ events at $m_H=125.5$ GeV
is 27\% (22\%) for $\ell=e$ and 33\% (27\%) for $\ell=\mu$ 
at $\sqrt{s}=8$ (7) TeV.
The relative efficiency is about 5\% higher in the VBF process and 5--10\%
lower in the $W$, $Z$, $t\bar t$-associated production modes, compared 
to signal events produced in the dominant gluon-fusion process. 
For $m_H$ increasing between 120 and 150 GeV the overall signal efficiency varies from $0.87$
to $1.25$ times the efficiency at $m_H=125.5$ GeV.

\subsection{Invariant-mass calculation}
\label{ss:resolution}
In order to improve the three-body invariant-mass resolution of the
Higgs boson candidate events and thus improve discrimination against
non-resonant background events, three corrections are applied to the
three-body mass $m_{\ell\ell\gamma}$.
First, the photon pseudorapidity $\eta^\gamma$ and its transverse
energy $\ET^\gamma = E^\gamma / \cosh \eta^\gamma$ are recalculated
using the identified primary vertex as the photon's origin, 
rather than the nominal interaction point 
(which is used in the standard ATLAS photon reconstruction).
Second, the muon momenta are corrected for collinear 
final-state-radiation (FSR)
by including any reconstructed electromagnetic cluster with $\ET$ above 1.5 GeV lying
close (typically with $\Delta R<0.15$) to a muon track.
Third, the lepton four-momenta are recomputed by means of a 
$Z$-mass-constrained kinematic fit previously used in the ATLAS $H\to4\ell$
search~\cite{ATLAS_Higgs}.
The photon direction and FSR corrections improve the invariant-mass 
resolution by about 1\% each, while the $Z$-mass constraint brings an
improvement of about 15--20\%.

Fig.~\ref{fig:signal_resolution_corrections}
illustrates the distributions of $m_{\mu\mu\gamma}$ and $m_{ee\gamma}$ 
for simulated signal events
from $gg\to H$ at $m_H=125$ GeV after all corrections.
The $m_{ee\gamma}$ resolution is about 8\% worse due to 
bremsstrahlung.
The $m_{\ell\ell\gamma}$ distribution is modelled with the sum 
of a Crystal Ball function (a Gaussian with a power-law tail),
representing the core of well-reconstructed events, 
and a small, wider Gaussian component describing the tails
of the distribution.
For $m_H=125.5$ GeV the typical mass resolution $\sigma_{CB}$ of the core
component of the $m_{\mu\mu\gamma}$ distribution is 1.6 GeV.

\subsection{Event classification}
\label{ss:categories}
The selected events are classified into four categories, based on 
the $pp$ centre-of-mass energy and the lepton flavour.
To enhance the sensitivity of the analysis, each event class is 
further divided into categories with different signal-to-background ratios and 
invariant-mass resolutions, based on (i) the
pseudorapidity difference $\Delta \eta_{Z\gamma}$ between the photon
and the $Z$ boson and (ii) $p_{\rm Tt}$,\footnote{$p_{\rm Tt} = |(\vec p_{\rm T}^\gamma + \vec p_{\rm T}^Z)\times \hat t|$ 
where $\hat t=(\vec{p}_{\rm T}^\gamma-\vec{p}_{\rm T}^Z)/|\vec{p}_{\rm T}^\gamma-\vec{p}_{\rm T}^Z|$
denotes the thrust axis in the transverse plane, and $\vec p_{\rm T}^\gamma$, 
$\vec p_{\rm T}^Z$ are the transverse momenta of the photon and the $Z$ boson.}
the component of the Higgs boson
candidate $\pT$ that is orthogonal to the $Z\gamma$ thrust axis 
in the transverse plane.
Signal events are typically characterised by a larger 
$p_{\rm Tt}$ and a smaller $\Delta \eta_{Z\gamma}$ 
compared to background events, which are mostly due
to $q\bar{q}\to Z+\gamma$ events in which the $Z$ boson
and the photon are back-to-back in the transverse plane.
Signal gluon-fusion events have on average smaller 
$p_{\rm Tt}$ and larger $\Delta \eta_{Z\gamma}$ than signal events 
in which the Higgs boson is produced either by VBF or in association 
with $W$, $Z$ or $t\bar{t}$ and thus is more boosted. 

Higgs boson candidates are classified as {\em high-} ({\em
  low-}) $p_{\rm Tt}$ candidates if their $p_{\rm Tt}$ is greater (smaller)
than 30 GeV. In the analysis of $\sqrt{s}=8$ TeV data, 
low-$p_{\rm Tt}$ candidates are further split into two classes, {\em
  high-} and {\em low-}$\Delta \eta_{Z\gamma}$, depending on whether 
$|\Delta \eta_{Z\gamma}|$ is greater or less than 2.0,
yielding a total of ten event categories.
\ifcountpages
\else
\begin{figure}[h]
  \begin{center}
    \includegraphics[width=\columnwidth]{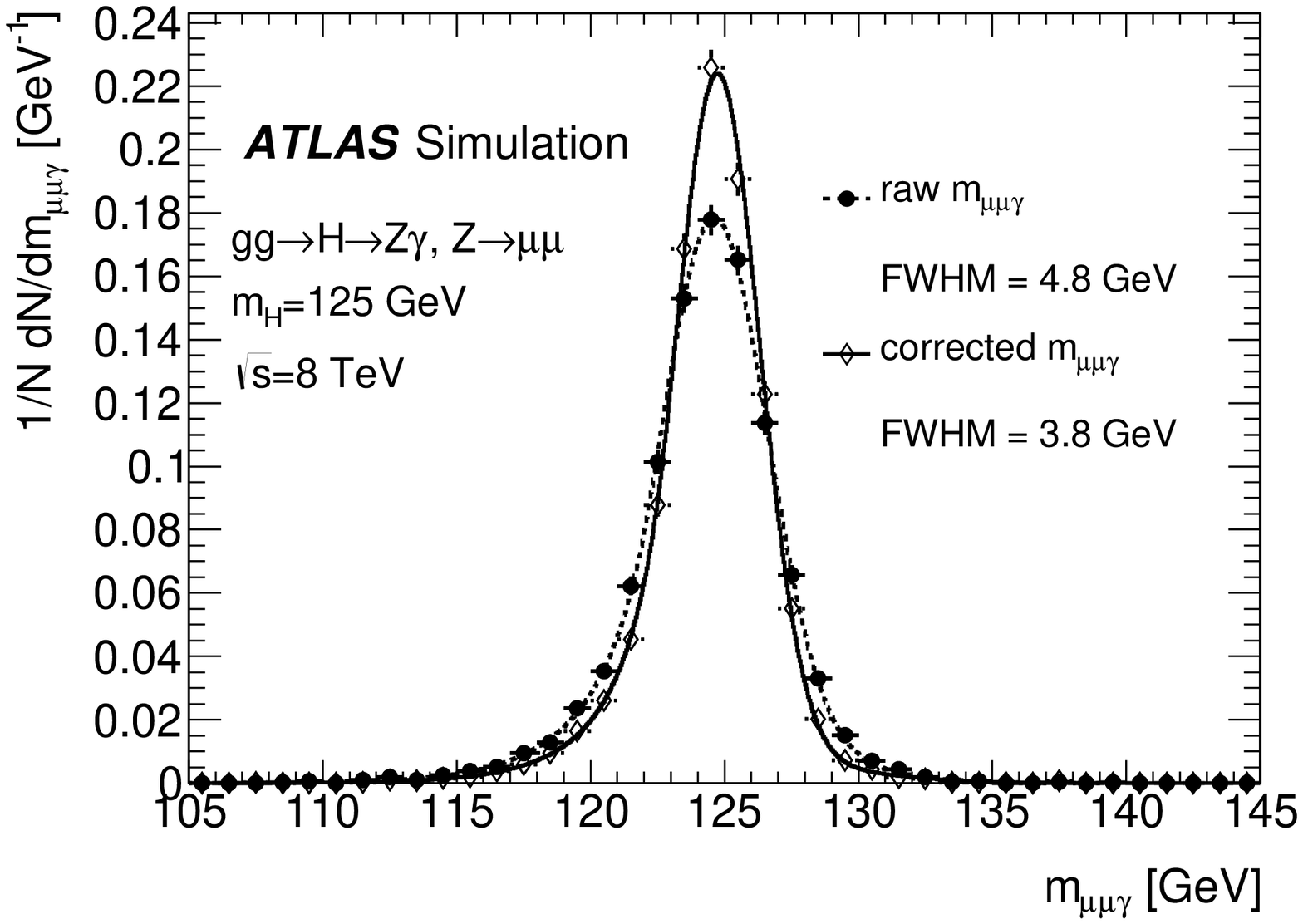}
    \includegraphics[width=\columnwidth]{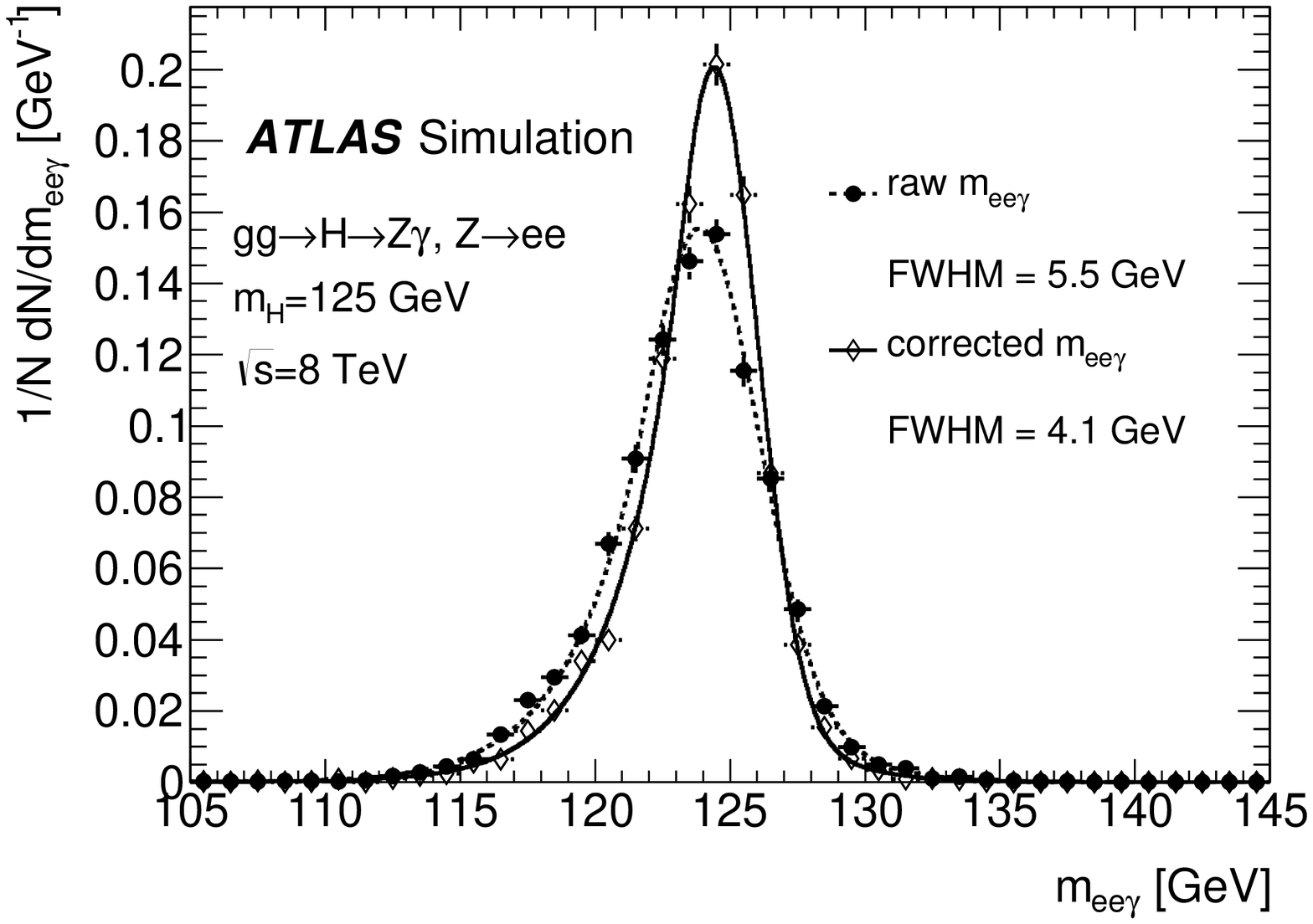}
    \caption{Three-body invariant-mass distribution for $H\to Z\gamma$, $Z\to\mu\mu$ (top)
      or $Z\to ee$ (bottom) selected events in the 8 TeV, $m_H=125$~GeV gluon-fusion signal simulation,
      after applying all analysis cuts, before (filled circles) and after (open diamonds)
      the corrections described in Section~\ref{ss:resolution}.
      The solid and dashed lines represent the fits of the points to the sum of a
      Crystal Ball and a Gaussian function.}
    \label{fig:signal_resolution_corrections}
  \end{center}
\end{figure}
\fi

As an example, the expected number of signal and background events in each category 
with invariant mass within a $\pm 5$~GeV window around $m_H = 125$ GeV,
the observed number of events in data in the same region,
and the full-width at half-maximum (FWHM) 
of the signal invariant-mass distribution,
are summarised in Table~\ref{tab:categories}.
Using this classification improves the signal sensitivity 
of this analysis by 33\% for a Higgs boson mass of 
125.5 GeV compared to a classification based only on the 
centre-of-mass energy and lepton flavour categories.
 
\begin{table}[h]
  \begin{small}
    \setlength{\tabcolsep}{4pt}
    \caption{Expected signal ($N_{\rm S}$) and background ($N_{\rm B}$) yields in
      a $\pm 5$~GeV mass window around $m_H=125$ GeV for each of the
      event categories under study.
      In addition, the observed number of events in data ($N_{\rm D}$) and
      the FWHM of the signal invariant-mass distribution, modelled as
      described in Section~\ref{ss:resolution}, are given. 
      The signal is assumed to have SM-like properties, including the
      production cross section times branching ratio.
      The background yield is extrapolated from the selected data event
      yield in the invariant-mass region outside the $\pm 5$~GeV
      window around $m_H=125$ GeV, using an analytic background model
      described in Section~\ref{sec:Results}.
      The uncertainty on the FWHM from the limited size of the
      simulated signal samples is negligible in comparison to the systematic
      uncertainties described in Section~\ref{sec:Systematics}.
    }
    \label{tab:categories}
    \begin{center}
      \begin{tabular}{lllcrrrrr}
        \hline\hline
        $\sqrt{s}$ & $\ell$ & Category & $N_{\rm S}$& $N_{\rm B}$ & $N_{\rm D}$ & $\frac{N_{\rm S}}{\sqrt{N_{\rm B}}}$ & FWHM  \\
        $[$TeV$]$  &        &        &      &       &       &                          & [GeV] \\
        \hline
        8 & $\mu$& high $p_{\rm Tt}$                      & 2.3 &  310 &  324 & 0.13 & 3.8 \\
        8 & $\mu$& low $p_{\rm Tt}$, low $\Delta \eta$    & 3.7 & 1600 & 1587 & 0.09 & 3.8 \\
        8 & $\mu$& low $p_{\rm Tt}$, high $\Delta \eta$   & 0.8 &  600 &  602 & 0.03 & 4.1 \\
        \hline
        8 & $e$  & high $p_{\rm Tt}$                      & 1.9 &  260 &  270 & 0.12 & 3.9 \\
        8 & $e$  & low $p_{\rm Tt}$, low $\Delta \eta$    & 2.9 & 1300 & 1304 & 0.08 & 4.2 \\
        8 & $e$  & low $p_{\rm Tt}$, high $\Delta \eta$   & 0.6 &  430 &  421 & 0.03 & 4.5 \\
        \hline
        7 & $\mu$& high $p_{\rm Tt}$                      & 0.4 &   40 &   40 & 0.06 & 3.9 \\
        7 & $\mu$& low $p_{\rm Tt}$                       & 0.6 &  340 &  335 & 0.03 & 3.9 \\
        \hline
        7 & $e$  & high $p_{\rm Tt}$                      & 0.3 &   25 &   21 & 0.06 & 3.9 \\
        7 & $e$  & low $p_{\rm Tt}$                       & 0.5 &  240 &  234 & 0.03 & 4.0 \\
        \hline\hline
      \end{tabular}
    \end{center}
  \end{small}
\end{table}

\subsection{Sample composition}
\label{ss:BkgEstimation}
The main backgrounds originate from continuum $Z$+$\gamma$, $Z\to\ell\ell$
production, from radiative $Z\to\ell\ell\gamma$ decays, and 
from $Z$+jet, $Z\to\ell\ell$ events in which a jet is misidentified 
as a photon.
Small contributions arise from $t\bar t$ and $WZ$ events.
Continuum $Z$+$\gamma$ events are either produced by $q\overline{q}$ 
in the $t$- or $u$- channels, 
or from parton-to-photon fragmentation.
The requirements $m_{\ell\ell}>m_Z-10$ GeV,
$m_{\ell\ell\gamma}>115$ GeV and $\Delta R_{\ell\gamma}>0.3$ 
suppress the contribution from $Z\to\ell\ell\gamma$, while
the photon isolation requirement reduces the importance of the
$Z$+$\gamma$ fragmentation component. The latter, together with the photon 
identification requirements, is also effective in reducing $Z$+jets 
events.

In this analysis, the estimated background composition is not used 
to determine the amount of expected background, which is directly 
fitted to the data mass spectrum, 
but is used to normalise the background Monte Carlo samples used
for the optimisation of the selection criteria and 
the choice of mass spectra background-fitting functions and 
the associated systematic uncertainties.
Since the amplitudes for $Z$+$\gamma$, $Z\to\ell\ell$ and $Z\to\ell\ell\gamma$
interfere, only the total $\ell\ell\gamma$ background from the sum of 
the two processes is considered, and denoted with $Z\gamma$ in the following.
A data-driven estimation of the background composition
is performed, based on a two-dimensional sideband method~\cite{Aad:2010sp,Aad:2013izg}
exploiting the distribution of the photon identification 
and isolation variables in control regions enriched in $Z$+jets events,
to estimate the relative $Z\gamma$ and $Z$+jets fractions in the 
selected sample.
The $Z\gamma$ and $Z$+jets contributions are estimated {\em in situ}
by applying this
technique to the data after subtracting the $1\%$ contribution 
from the $t\bar t$ and $WZ$ backgrounds. 
Simulated events are used to estimate the small backgrounds from 
$t\bar t$ and $WZ$ production (normalised to the data luminosity using the NLO
MC cross sections), on which a conservative uncertainty of $\pm 50\%$
accounts for observed data-MC differences 
in the rates of fake photons and leptons from misidentified jets 
as well as for the uncertainties
on the MC cross section due to the missing higher orders of the 
perturbative expansion and the PDF uncertainties.
Simulated events are also used to determine
the $Z\gamma$ contamination in the
$Z$+jet background control regions and the correlation between 
photon identification and photon isolation for $Z$+jet events.
The contribution to the control regions from the $H\to Z\gamma$ signal is expected to be 
small compared to the background and is neglected in this
study.
The fractions of $Z\gamma$, $Z$+jets and other ($t\bar{t}+WZ$)
backgrounds are estimated to be around 
82\%, 17\% and 1\% at both $\sqrt{s}=7$ and 8 TeV.
The relative uncertainty on the $Z\gamma$ purity is around 5\%, 
dominated by the uncertainty on the correlation between the 
photon identification and isolation in $Z$+jet events, which is
estimated by comparing the ALPGEN and SHERPA predictions.
Good agreement between data and simulation is observed in the 
distributions of $m_{\ell\ell\gamma}$, as well as in the 
distributions of several other kinematic quantities that were also 
studied, including the dilepton invariant mass and the lepton
and photon transverse momenta, pseudorapidity and azimuth.

\section{Experimental systematic uncertainties}
\label{sec:Systematics}

The following sources of experimental systematic uncertainties 
on the expected signal yields in each category were considered:
\begin{itemize}
\item
The luminosity uncertainty is 1.8\% for the 2011 
data~\cite{lumi2011} and 2.8\% for the 2012 
data.\footnote{The luminosity of the 2012 data is derived, following the same
methodology as that detailed in Ref.~\cite{lumi2011}, 
from a preliminary calibration of the luminosity scale 
derived from beam-separation scans performed in November 2012.} 

\item 
The uncertainty from the photon identification efficiency
is obtained from a comparison between data-driven measurements 
and the simulated efficiencies in various photon and electron 
control samples~\cite{ATLAS-CONF-2012-123} 
and varies between 2.6\% and 3.1\% depending on the category.
The uncertainty from the photon reconstruction efficiency is negligible 
compared to that from the identification efficiency.

\item
The uncertainty from the electron trigger, reconstruction and
identification efficiencies is estimated by varying the efficiency
corrections applied to the simulation within the uncertainties of
data-driven efficiency measurements. 
The total uncertainty, for events in which the $Z$ boson candidate 
decays to electrons, varies between 2.5\% and 3\% depending on the 
category.
The lepton reconstruction, identification and trigger efficiencies, 
as well as their energy and momentum scales and resolutions, 
are determined using large control samples of  
$Z\to\ell\ell$, $W\to\ell\nu$ and $J/\psi \to \ell\ell$
events~\cite{muonreco,Aad:2011mk}.

\end{itemize}
Other sources of uncertainty (muon trigger, reconstruction and
identification efficiencies, lepton energy scale, resolution,
and impact parameter selection efficiencies, lepton and photon
isolation efficiencies)
were investigated and found to have a negligible impact on the
signal yield compared to the mentioned sources of uncertainty.
The total relative uncertainty on the signal efficiency in each category 
is less than 5\%, more than twice as small as the corresponding
theoretical systematic uncertainty on the SM production
cross section times branching ratio, described in Section~\ref{sec:SimSamples}.
The uncertainty in the population of the $p_{\rm Tt}$ categories due
to the description of the Higgs boson $\pT$ spectrum is
determined by varying the QCD scales and PDFs used
in the HRES2 program. It is estimated to vary between 1.8\% and 3.6\%
depending on the category.

The following sources of experimental systematic uncertainties 
on the signal $m_{\ell\ell\gamma}$ distribution were considered:
\begin{itemize}
\item The uncertainty on the peak position (0.2 GeV) is dominated by the photon 
    energy scale uncertainty, which arises from the following
    sources: the calibration of the electron energy scale from $Z\to
    ee$ events, the uncertainty on its extrapolation to the energy
    scale of photons, dominated by the description of the detector
    material, and imperfect knowledge of the energy scale of the presampler
    detector located in front of the electromagnetic calorimeter.  
 \item The uncertainty from the photon and electron energy resolution
  is estimated as the relative variation of the width of
  the signal $m_{\ell\ell\gamma}$ distribution after
  varying the corrections to the resolution 
  of the electromagnetic particle response in the simulation
  within their uncertainties.
  It amounts to 3\% for events in which the $Z$ boson 
  candidate decays to muons and to 10\% for events in which 
  the $Z$ boson candidate decays to electrons. 
\item The uncertainty from the muon momentum resolution  
  is estimated as the relative variation of the width of
  the signal $m_{\ell\ell\gamma}$ distribution after
  varying the muon momentum smearing corrections
  within their uncertainties. It is smaller than 1.5\%.
\end{itemize}

To extract the signal, the background is estimated from the observed
$m_{\ell\ell\gamma}$ distribution by assuming an analytical model,
chosen from several alternatives to provide the best sensitivity to
the signal while limiting the possible bias in the fitted signal 
to be within $\pm 20\%$ of the statistical uncertainty on the signal
yield due to background fluctuations.
The models are tested by performing signal+background fits
of the $m_{\ell\ell\gamma}$ distribution of large simulated
background-only samples scaled to the luminosity of the data
and evaluating the ratio of the fitted
signal yield to the statistical uncertainty on the fitted signal
itself.
The largest observed bias in the fitted signal for any Higgs boson mass
in the range 120--150 GeV is taken as an additional systematic
uncertainty; it varies between 0.5 events in poorly populated categories
and 8.3 events in highly populated ones.

All systematic uncertainties, except that on the luminosity, are taken as 
fully correlated between the $\sqrt{s}=7$ TeV and the $\sqrt{s}=8$ TeV analyses.

\section{Results}
\label{sec:Results}

\subsection{Likelihood function}
\label{ss:likelihood}
The final discrimination between signal and background events is 
based on a likelihood fit to the $m_{\ell\ell\gamma}$ spectra
in the invariant-mass region $115<m_{\ell\ell\gamma}<170$~GeV.
The likelihood function depends on a single parameter of interest,
the Higgs boson production signal strength $\mu$, defined as the 
signal yield normalised to the SM expectation, as well as on 
several nuisance parameters that describe
the shape and normalisation of the background 
distribution in each event category and the systematic uncertainties.
Results for the inclusive cross section times branching ratio 
are also provided. 
In that case, the likelihood function depends on two parameters of 
interest, the signal cross sections times branching ratios at 
$\sqrt{s}=7$ TeV and $\sqrt{s}=8$ TeV, and the 
systematic uncertainties on the SM cross sections
and branching ratios.

The background model in each event category is chosen based on the 
studies of sensitivity versus bias described in the previous section.
For 2012 data, fifth- and fourth-order polynomials are chosen to model the
background in the 
low-$p_{\rm Tt}$
categories while an exponentiated second-order polynomial is chosen
for the high-$p_{\rm Tt}$ categories.
For 2011 data, a fourth-order polynomial is used for the 
low-$p_{\rm Tt}$ categories and an exponential function is chosen for the
high-$p_{\rm Tt}$ ones.
The signal resolution functions in each category are described by the
model illustrated in Section~\ref{ss:resolution}, fixing the fraction of
events in each category to the MC predictions. 
For each fixed value of the Higgs boson mass between 120 and 150 GeV, 
in steps of 0.5 GeV, the parameters of the signal model 
are obtained, separately for each event category, through interpolation 
of the fully simulated MC samples.

For each of the nuisance parameters describing systematic uncertainties the 
likelihood is multiplied by a constraint term for each of
the experimental systematic uncertainties evaluated as described in Section~\ref{sec:Systematics}.
For systematic uncertainties affecting the expected total signal yields
for different centre-of-mass or lepton flavour, a log-normal constraint 
is used 
while for the uncertainties on the fractions of signal events in different 
$p_{\rm {Tt}}-\Delta\eta_{Z\gamma}$ categories and on the 
signal $m_{\ell\ell\gamma}$ resolution a Gaussian constraint is 
used~\cite{HiggsCombStatMethod2011}. 

\subsection{Statistical analysis}
The data are compared to background and signal-plus-background
hypotheses using a profile likelihood test 
statistic~\cite{HiggsCombStatMethod2011}.
Higgs boson decays to final states other than $\ell\ell\gamma$ 
are expected to contribute negligibly to the background in the selected 
sample.
For each fixed value of the Higgs boson mass between 120 and 150 GeV 
fits are performed in steps of 0.5 GeV to determine the best value of
$\mu$ ($\hat \mu$) or to maximise the likelihood with respect to all
the nuisance parameters for alternative values of $\mu$, 
including $\mu=0$ (background-only
hypothesis) and $\mu=1$ (background plus Higgs boson of that mass,
with SM-like production cross section times branching ratio).
The compatibility between the data and the background-only hypothesis
is quantified by the $p$-value of the $\mu=0$ hypothesis, $p_0$, which
provides an estimate of the significance of a possible observation.
Upper limits on the signal strength at 95\% $CL$ are set 
using a modified frequentist ($CL_{s}$) method~\cite{cls}, by
identifying the value $\mu_{\rm up}$ for which the $CL_s$ is
equal to $0.05$.
Closed-form asymptotic formulae~\cite{Cowan:2010js} are used to derive
the results.
Fits to the data are performed to obtain observed results. 
Fits to Asimov pseudo-data~\cite{Cowan:2010js}, generated either according to
the $\mu=1$ or $\mu=0$ hypotheses, are performed to compute
expected $p_0$ and $CL_s$ upper limits, respectively.

\ifcountpages
\else
 \begin{figure}[!htbp]
  \begin{center}
    \includegraphics[width=\columnwidth]{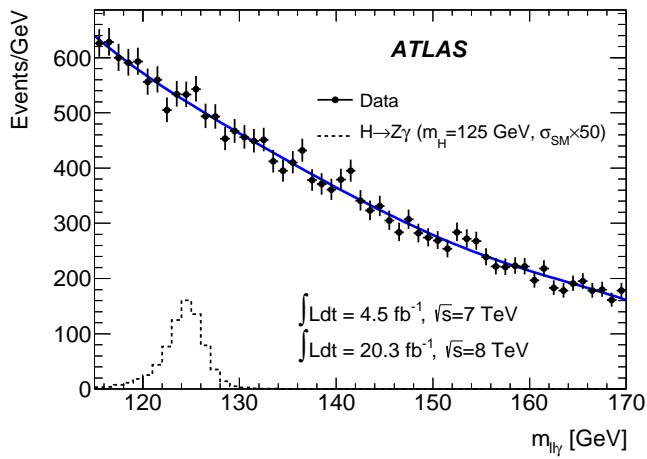}
    \caption{Distribution of the reconstructed $\ell\ell\gamma$ invariant 
      mass in data, after combining all the event categories (points with error bars).
      The solid blue line shows the sum of background-only fits to the data performed 
      in each category.
      The dashed histogram corresponds to the signal expectation
      for a Higgs boson mass of 125 GeV decaying to $Z\gamma$ at 50 times
      the SM-predicted rate.
    }
    \label{fig:Mllg_data_bkgonly_fit}
  \end{center}
\end{figure}
\fi
Figure~\ref{fig:Mllg_data_bkgonly_fit} shows the $m_{\ell\ell\gamma}$ 
distribution of all events selected in data, compared to the sum of 
the background-only fits to the data in each of the ten event categories.
No significant excess with respect to the background is visible, 
and the observed $p_0$ is compatible
with the data being composed of background only. 
The smallest $p_0$ (\pvaluelow), corresponding to a significance
of \sigmahigh~$\sigma$, occurs for a mass of 141 GeV.
The expected $p_0$ ranges between \pvaluelowexp\ and
\pvaluehighexp\ for a Higgs boson with a mass $120<m_H<150$~GeV and
SM-like cross section and branching ratio,
corresponding to significances
around \sigmalowexp~$\sigma$. 
The expected  $p_0$ at $m_H=125.5$~GeV is \pvaluehiggsexp, 
corresponding to a significance of \sigmahiggsexp\ $\sigma$, 
while the observed $p_0$ at the same mass is
\pvaluehiggs\ (\sigmahiggs\ $\sigma$).

Observed and expected $95\%$ $CL$ upper limits on the
value of the signal strength $\mu$ are derived and 
shown in Fig.~\ref{fig:ExpectedExclusion_1}.
The expected limit ranges between 
\excllowexp\ and \exclhighexp\ and the observed limit varies
between \excllow\ and \exclhigh\
for a Higgs boson mass between 120 and 150 GeV.
In particular, for a mass of 125.5 GeV,
the observed and expected limits are equal 
to \exclhiggs\ and \exclhiggsexp\ times the Standard Model prediction,
respectively.
At the same mass the expected limit on $\mu$ assuming the existence
of a SM ($\mu=1$) Higgs boson with $m_H=125.5$ GeV is 10.
The results are dominated by the statistical uncertainties:
neglecting all systematic uncertainties, the observed and expected
95\% $CL$ limits on the cross section at 125.5 GeV 
decrease by about 5\%.

\ifcountpages
\else
\begin{figure}[!htbp]
\centering
\includegraphics[width=\columnwidth]{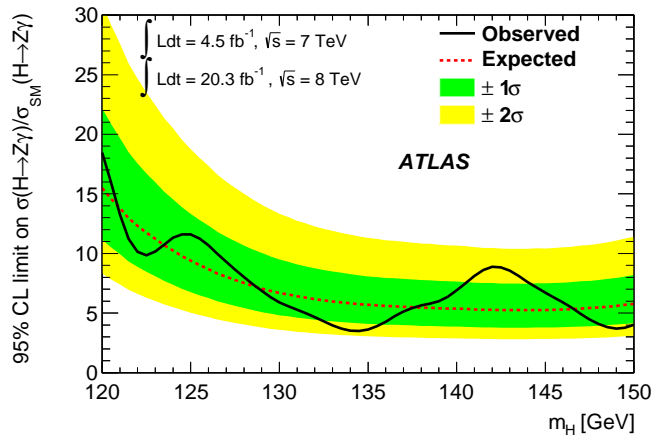}
\caption{Observed 95\% $CL$ limits (solid black line) 
  on the production cross section of a SM Higgs boson 
  decaying to $Z\gamma$ divided by the SM expectation.
  The limits are computed as a function of the Higgs boson 
  mass.
  The median expected 95\% $CL$ exclusion limits (dashed red line),
  in the case of no expected signal, are also shown. 
  The green and yellow bands correspond to the $\pm 1\sigma$ 
  and $\pm2\sigma$ intervals.
}
\label{fig:ExpectedExclusion_1}
\end{figure}
\fi

Upper limits on the $pp\to H\to Z\gamma$ cross section times 
branching ratio are also derived at 95\% $CL$, for $\sqrt{s}=7$ and 8 TeV.
For $\sqrt{s}=8$ TeV, the limit ranges between 0.13 and 0.5 pb; for 
$\sqrt{s}=7$ TeV, it ranges between 0.20 and 0.8 pb.

\section{Conclusions}
A search for a Higgs boson in the decay channel $H\to
Z\gamma$, $Z\to\ell\ell$ ($\ell=e,\mu$), 
in the mass range 120-150 GeV, was performed using
\lumiseventev~\ifb\ of proton--proton collisions at $\sqrt{s}=7$~TeV and
\lumieighttev~\ifb\ of proton--proton collisions at $\sqrt{s}=8$~TeV
recorded with the ATLAS detector at the LHC.
No excess with respect to the background is found in the $\ell\ell\gamma$
invariant-mass distribution and 95\% $CL$
upper limits on the cross section times branching ratio are derived.
For $\sqrt{s}=8$ TeV, the limit ranges between 0.13 and 0.5 pb.
Combining $\sqrt{s}=7$ and 8 TeV data and dividing the cross section by
the Standard Model expectation, for a mass of 125.5 GeV,
the observed 95\% confidence limit is \exclhiggs\ 
times the SM prediction.

\section*{Acknowledgments}


We thank CERN for the very successful operation of the LHC, as well as the
support staff from our institutions without whom ATLAS could not be
operated efficiently.

We acknowledge the support of ANPCyT, Argentina; YerPhI, Armenia; ARC,
Australia; BMWF and FWF, Austria; ANAS, Azerbaijan; SSTC, Belarus; CNPq and FAPESP,
Brazil; NSERC, NRC and CFI, Canada; CERN; CONICYT, Chile; CAS, MOST and NSFC,
China; COLCIENCIAS, Colombia; MSMT CR, MPO CR and VSC CR, Czech Republic;
DNRF, DNSRC and Lundbeck Foundation, Denmark; EPLANET, ERC and NSRF, European Union;
IN2P3-CNRS, CEA-DSM/IRFU, France; GNSF, Georgia; BMBF, DFG, HGF, MPG and AvH
Foundation, Germany; GSRT and NSRF, Greece; ISF, MINERVA, GIF, DIP and Benoziyo Center,
Israel; INFN, Italy; MEXT and JSPS, Japan; CNRST, Morocco; FOM and NWO,
Netherlands; BRF and RCN, Norway; MNiSW and NCN, Poland; GRICES and FCT, Portugal; MNE/IFA, Romania; MES of Russia and ROSATOM, Russian Federation; JINR; MSTD,
Serbia; MSSR, Slovakia; ARRS and MIZ\v{S}, Slovenia; DST/NRF, South Africa;
MINECO, Spain; SRC and Wallenberg Foundation, Sweden; SER, SNSF and Cantons of
Bern and Geneva, Switzerland; NSC, Taiwan; TAEK, Turkey; STFC, the Royal
Society and Leverhulme Trust, United Kingdom; DOE and NSF, United States of
America.

The crucial computing support from all WLCG partners is acknowledged
gratefully, in particular from CERN and the ATLAS Tier-1 facilities at
TRIUMF (Canada), NDGF (Denmark, Norway, Sweden), CC-IN2P3 (France),
KIT/GridKA (Germany), INFN-CNAF (Italy), NL-T1 (Netherlands), PIC (Spain),
ASGC (Taiwan), RAL (UK) and BNL (USA) and in the Tier-2 facilities
worldwide.

\section*{References}
\bibliographystyle{elsarticle-num}
\bibliography{paper}

\begin{thebibliography}{10}
\expandafter\ifx\csname url\endcsname\relax
  \def\url#1{\texttt{#1}}\fi
\expandafter\ifx\csname urlprefix\endcsname\relax\def\urlprefix{URL }\fi
\expandafter\ifx\csname href\endcsname\relax
  \def\href#1#2{#2} \def\path#1{#1}\fi

\bibitem{ATLAS_Higgs}
{ATLAS Collaboration}, {Observation of a new particle in the search for the
  Standard Model Higgs boson with the ATLAS detector at the LHC}, Phys.~Lett.~B
  716 (2012) 1.
\newblock \href {http://arxiv.org/abs/1207.7214} {\path{arXiv:1207.7214}},
  \href {http://dx.doi.org/10.1016/j.physletb.2012.08.020}
  {\path{doi:10.1016/j.physletb.2012.08.020}}.

\bibitem{CMS_Higgs}
{CMS Collaboration}, {Observation of a new boson at a mass of 125 GeV with the
  CMS experiment at the LHC}, Phys.~Lett.~B 716 (2012) 30.
\newblock \href {http://arxiv.org/abs/1207.7235} {\path{arXiv:1207.7235}},
  \href {http://dx.doi.org/10.1016/j.physletb.2012.08.021}
  {\path{doi:10.1016/j.physletb.2012.08.021}}.

\bibitem{ATLAS_couplings}
{ATLAS Collaboration}, {Measurements of Higgs boson production and couplings in
  diboson final states with the ATLAS detector at the LHC}, Phys.~Lett. B 726
  (2013) 88--119.
\newblock \href {http://arxiv.org/abs/1307.1427} {\path{arXiv:1307.1427}},
  \href {http://dx.doi.org/10.1016/j.physletb.2013.08.010}
  {\path{doi:10.1016/j.physletb.2013.08.010}}.

\bibitem{CMS_couplings}
{CMS Collaboration}, {Observation of a new boson with mass near 125 GeV in $pp$
  collisions at $\sqrt{s}$ = 7 and 8 TeV}, JHEP 1306 (2013) 081.
\newblock \href {http://arxiv.org/abs/1303.4571} {\path{arXiv:1303.4571}},
  \href {http://dx.doi.org/10.1007/JHEP06(2013)081}
  {\path{doi:10.1007/JHEP06(2013)081}}.

\bibitem{ATLAS_spin}
{ATLAS Collaboration}, {Evidence for the spin-0 nature of the Higgs boson using
  ATLAS data}, Phys.~Lett. B 726 (2013) 120--144.
\newblock \href {http://arxiv.org/abs/1307.1432} {\path{arXiv:1307.1432}},
  \href {http://dx.doi.org/10.1016/j.physletb.2013.08.026}
  {\path{doi:10.1016/j.physletb.2013.08.026}}.

\bibitem{CMS_spin}
{CMS Collaboration}, {Study of the mass and spin-parity of the Higgs boson
  candidate via its decays to Z boson pairs}, Phys.~Rev.~Lett. 110 (2013)
  081803.
\newblock \href {http://arxiv.org/abs/1212.6639} {\path{arXiv:1212.6639}}.

\bibitem{CMS-HZG}
{CMS Collaboration}, {Search for a Higgs boson decaying into a $Z$ and a photon
  in $pp$ collisions at $\sqrt{s}$ = 7 and 8 TeV}, Phys.~Lett. B 726 (2013)
  587--609.
\newblock \href {http://arxiv.org/abs/1307.5515} {\path{arXiv:1307.5515}},
  \href {http://dx.doi.org/10.1016/j.physletb.2013.09.057}
  {\path{doi:10.1016/j.physletb.2013.09.057}}.

\bibitem{LHCHiggsCrossSectionWorkingGroup:2011ti}
{LHC Higgs Cross Section Working Group}, S.~Dittmaier, C.~Mariotti,
  G.~Passarino, R.~Tanaka~(Eds.), {Handbook of LHC Higgs cross sections: 1.
  Inclusive observables}, {CERN-2011-002. }\href
  {http://arxiv.org/abs/1101.0593} {\path{arXiv:1101.0593}}.

\bibitem{LHCHiggsCrossSectionWorkingGroup:2012vm}
{LHC Higgs Cross Section Working Group}, S.~Dittmaier, C.~Mariotti,
  G.~Passarino, R.~Tanaka~(Eds.), {Handbook of LHC Higgs cross sections: 2.
  Differential distributions}, {CERN-2012-002. }\href
  {http://arxiv.org/abs/1201.3084} {\path{arXiv:1201.3084}}.

\bibitem{Heinemeyer:2013tqa}
{LHC Higgs Cross Section Working Group}, S.~Heinemeyer, C.~Mariotti,
  G.~Passarino, R.~Tanaka~(Eds.), {Handbook of LHC Higgs Cross Sections: 3.
  Higgs Properties}, {CERN-2013-004. }\href {http://arxiv.org/abs/1307.1347}
  {\path{arXiv:1307.1347}}.

\bibitem{low}
I.~Low, J.~Lykken, G.~Shaughnessy, {Singlet scalars as Higgs imposters at the
  Large Hadron Collider}, Phys.~Rev. D 84 (2011) 035027.
\newblock \href {http://arxiv.org/abs/1105.4587} {\path{arXiv:1105.4587}},
  \href {http://dx.doi.org/10.1103/PhysRevD.84.035027}
  {\path{doi:10.1103/PhysRevD.84.035027}}.

\bibitem{low2}
I.~Low, J.~Lykken, G.~Shaughnessy, {Have we observed the Higgs (imposter)?},
  Phys.~Rev. D 86 (2012) 093012.
\newblock \href {http://arxiv.org/abs/1207.1093} {\path{arXiv:1207.1093}},
  \href {http://dx.doi.org/10.1103/PhysRevD.86.093012}
  {\path{doi:10.1103/PhysRevD.86.093012}}.

\bibitem{Azatov:2013ura}
A.~Azatov, R.~Contino, A.~Di~Iura, J.~Galloway, {New Prospects for Higgs
  Compositeness in $h\to Z \gamma$}, Phys.~Rev. D 88 (2013) 075019.
\newblock \href {http://arxiv.org/abs/1308.2676} {\path{arXiv:1308.2676}},
  \href {http://dx.doi.org/10.1103/PhysRevD.88.075019}
  {\path{doi:10.1103/PhysRevD.88.075019}}.

\bibitem{extensions}
C.-W. Chiang, K.~Yagyu, {Higgs boson decays to $\gamma\gamma$ and $Z \gamma$ in
  models with Higgs extensions}, Phys.~Rev. D 87 (2013) 033003.
\newblock \href {http://arxiv.org/abs/1207.1065} {\path{arXiv:1207.1065}},
  \href {http://dx.doi.org/10.1103/PhysRevD.87.033003}
  {\path{doi:10.1103/PhysRevD.87.033003}}.

\bibitem{Chen:2013vi}
C.-S. Chen, C.-Q. Geng, D.~Huang, L.-H. Tsai, {New Scalar Contributions to
  $h\to Z\gamma$}, Phys.~Rev. D 87 (2013) 075019.
\newblock \href {http://arxiv.org/abs/1301.4694} {\path{arXiv:1301.4694}},
  \href {http://dx.doi.org/10.1103/PhysRevD.87.075019}
  {\path{doi:10.1103/PhysRevD.87.075019}}.

\bibitem{carena}
M.~Carena, I.~Low, C.~E. Wagner, {Implications of a Modified Higgs to Diphoton
  Decay Width}, JHEP 1208 (2012) 060.
\newblock \href {http://arxiv.org/abs/1206.1082} {\path{arXiv:1206.1082}},
  \href {http://dx.doi.org/10.1007/JHEP08(2012)060}
  {\path{doi:10.1007/JHEP08(2012)060}}.

\bibitem{Aad:2008zzm}
{ATLAS Collaboration}, {The ATLAS experiment at the CERN Large Hadron
  Collider}, JINST 3 (2008) S08003.
\newblock \href {http://dx.doi.org/10.1088/1748-0221/3/08/S08003}
  {\path{doi:10.1088/1748-0221/3/08/S08003}}.

\bibitem{Trigger}
{ATLAS Collaboration}, {Performance of the ATLAS trigger system in 2010},
  Eur.~Phys.~J. C 72 (2012) 1849.
\newblock \href {http://arxiv.org/abs/1110.1530} {\path{arXiv:1110.1530}},
  \href {http://dx.doi.org/10.1140/epjc/s10052-011-1849-1}
  {\path{doi:10.1140/epjc/s10052-011-1849-1}}.

\bibitem{lumi2011}
{ATLAS Collaboration}, {Improved luminosity determination in $pp$ collisions at
  $\sqrt{s} = 7$ TeV using the ATLAS detector at the LHC}, Eur.~Phys.~J. C 73
  (2013) 2518.
\newblock \href {http://arxiv.org/abs/1302.4393} {\path{arXiv:1302.4393}},
  \href {http://dx.doi.org/10.1140/epjc/s10052-013-2518-3}
  {\path{doi:10.1140/epjc/s10052-013-2518-3}}.

\bibitem{powheg_ggF}
{S. Alioli, P. Nason, C. Oleari and E. Re}, {NLO Higgs boson production via
  gluon fusion matched with shower in POWHEG}, JHEP 0904 (2009) 002.
\newblock \href {http://arxiv.org/abs/0812.0578} {\path{arXiv:0812.0578}},
  \href {http://dx.doi.org/10.1088/1126-6708/2009/04/002}
  {\path{doi:10.1088/1126-6708/2009/04/002}}.

\bibitem{powheg_VBF}
{P. Nason and C. Oleari}, {NLO Higgs boson production via vector-boson fusion
  matched with shower in POWHEG}, JHEP 1002 (2010) 037.
\newblock \href {http://arxiv.org/abs/0911.5299} {\path{arXiv:0911.5299}},
  \href {http://dx.doi.org/10.1007/JHEP02(2010)037}
  {\path{doi:10.1007/JHEP02(2010)037}}.

\bibitem{Bagnaschi:2011tu}
E.~Bagnaschi, G.~Degrassi, P.~Slavich, A.~Vicini, {Higgs production via gluon
  fusion in the POWHEG approach in the SM and in the MSSM}, JHEP 1202 (2012)
  088.
\newblock \href {http://arxiv.org/abs/1111.2854} {\path{arXiv:1111.2854}},
  \href {http://dx.doi.org/10.1007/JHEP02(2012)088}
  {\path{doi:10.1007/JHEP02(2012)088}}.

\bibitem{Pythia8}
{T. Sj\"{o}strand, S. Mrenna, P. Skands}, {A brief introduction to PYTHIA 8.1},
  Comput.~Phys.~Commun. 178 (2008) 852--867.
\newblock \href {http://arxiv.org/abs/0710.3820} {\path{arXiv:0710.3820}},
  \href {http://dx.doi.org/10.1016/j.cpc.2008.01.036}
  {\path{doi:10.1016/j.cpc.2008.01.036}}.

\bibitem{sherpa2}
T.~Gleisberg, et~al., {Event generation with SHERPA 1.1}, JHEP 0902 (2009) 007.
\newblock \href {http://arxiv.org/abs/0811.4622} {\path{arXiv:0811.4622}},
  \href {http://dx.doi.org/10.1088/1126-6708/2009/02/007}
  {\path{doi:10.1088/1126-6708/2009/02/007}}.

\bibitem{sherpa3}
S.~Hoeche, S.~Schumann, F.~Siegert, {Hard photon production and matrix-element
  parton-shower merging}, Phys.~Rev.~D 81 (2010) 034026.
\newblock \href {http://arxiv.org/abs/0912.3501} {\path{arXiv:0912.3501}},
  \href {http://dx.doi.org/10.1103/PhysRevD.81.034026}
  {\path{doi:10.1103/PhysRevD.81.034026}}.

\bibitem{Mangano:2002ea}
M.~L. Mangano, et~al., {ALPGEN, a generator for hard multiparton processes in
  hadronic collisions}, JHEP 0307 (2003) 001.
\newblock \href {http://arxiv.org/abs/hep-ph/0206293}
  {\path{arXiv:hep-ph/0206293}}, \href
  {http://dx.doi.org/10.1088/1126-6708/2003/07/001}
  {\path{doi:10.1088/1126-6708/2003/07/001}}.

\bibitem{Herwig}
G.~Corcella, et~al., {HERWIG 6: an event generator for hadron emission
  reactions with interfering gluons (including super-symmetric processes) },
  JHEP 0101 (2001) 010.
\newblock \href {http://dx.doi.org/10.1088/1126-6708/2001/01/010}
  {\path{doi:10.1088/1126-6708/2001/01/010}}.

\bibitem{Frixione:2002ik}
S.~Frixione, B.~R. Webber, {Matching NLO QCD computations and parton shower
  simulations}, JHEP 0206 (2002) 029.
\newblock \href {http://arxiv.org/abs/hep-ph/0204244}
  {\path{arXiv:hep-ph/0204244}}.

\bibitem{Frixione:2003ei}
S.~Frixione, P.~Nason, B.~R. Webber, {Matching NLO QCD and parton showers in
  heavy flavour production}, JHEP 0308 (2003) 007.
\newblock \href {http://arxiv.org/abs/hep-ph/0305252}
  {\path{arXiv:hep-ph/0305252}}, \href
  {http://dx.doi.org/10.1088/1126-6708/2003/08/007}
  {\path{doi:10.1088/1126-6708/2003/08/007}}.

\bibitem{Lai:2010vv}
H.-L. Lai, M.~Guzzi, J.~Huston, Z.~Li, P.~M. Nadolsky, et~al., {New parton
  distributions for collider physics}, Phys.~Rev.~D 82 (2010) 074024.
\newblock \href {http://arxiv.org/abs/1007.2241} {\path{arXiv:1007.2241}},
  \href {http://dx.doi.org/10.1103/PhysRevD.82.074024}
  {\path{doi:10.1103/PhysRevD.82.074024}}.

\bibitem{HRES2}
M.~Grazzini, H.~Sargsyan, {Heavy-quark mass effects in Higgs boson production
  at the LHC}, JHEP 1309 (2013) 129.
\newblock \href {http://arxiv.org/abs/1306.4581} {\path{arXiv:1306.4581}},
  \href {http://dx.doi.org/10.1007/JHEP09(2013)129}
  {\path{doi:10.1007/JHEP09(2013)129}}.

\bibitem{JHEP_0207_012}
J.~Pumplin, et~al., New generation of parton distributions with uncertainties
  from global qcd analysis, JHEP 0207 (2002) 012.
\newblock \href {http://dx.doi.org/10.1088/1126-6708/2002/07/012}
  {\path{doi:10.1088/1126-6708/2002/07/012}}.

\bibitem{Harlander:2002wh}
R.~V. Harlander, W.~B. Kilgore, {Next-to-next-to-leading order Higgs production
  at hadron colliders}, Phys. Rev. Lett. 88 (2002) 201801.
\newblock \href {http://arxiv.org/abs/hep-ph/0201206}
  {\path{arXiv:hep-ph/0201206}}, \href
  {http://dx.doi.org/10.1103/PhysRevLett.88.201801}
  {\path{doi:10.1103/PhysRevLett.88.201801}}.

\bibitem{Anastasiou:2002yz}
C.~Anastasiou, K.~Melnikov, {Higgs boson production at hadron colliders in NNLO
  QCD}, Nucl. Phys. B~646 (2002) 220--256.
\newblock \href {http://arxiv.org/abs/hep-ph/0207004}
  {\path{arXiv:hep-ph/0207004}}, \href
  {http://dx.doi.org/10.1016/S0550-3213(02)00837-4}
  {\path{doi:10.1016/S0550-3213(02)00837-4}}.

\bibitem{Ravindran:2003um}
V.~Ravindran, J.~Smith, W.~L. van Neerven, {NNLO corrections to the total cross
  section for Higgs boson production in hadron hadron collisions}, Nucl. Phys.
  B~665 (2003) 325--366.
\newblock \href {http://arxiv.org/abs/hep-ph/0302135}
  {\path{arXiv:hep-ph/0302135}}, \href
  {http://dx.doi.org/10.1016/S0550-3213(03)00457-7}
  {\path{doi:10.1016/S0550-3213(03)00457-7}}.

\bibitem{Anastasiou:2012hx}
C.~Anastasiou, S.~Buehler, F.~Herzog, A.~Lazopoulos, {Inclusive Higgs boson
  cross-section for the LHC at 8 TeV}, JHEP 1204 (2012) 004.
\newblock \href {http://arxiv.org/abs/1202.3638} {\path{arXiv:1202.3638}},
  \href {http://dx.doi.org/10.1007/JHEP04(2012)004}
  {\path{doi:10.1007/JHEP04(2012)004}}.

\bibitem{deFlorian:2012yg}
D.~de~Florian, M.~Grazzini, {Higgs production at the LHC: updated cross
  sections at $\sqrt{s}=8$ TeV}, Phys.~Lett. B 718 (2012) 117--120.
\newblock \href {http://arxiv.org/abs/1206.4133} {\path{arXiv:1206.4133}},
  \href {http://dx.doi.org/10.1016/j.physletb.2012.10.019}
  {\path{doi:10.1016/j.physletb.2012.10.019}}.

\bibitem{Aglietti:2004nj}
U.~Aglietti, R.~Bonciani, G.~Degrassi, A.~Vicini, {Two-loop light fermion
  contribution to Higgs production and decays}, Phys.~Lett. B~595 (2004)
  432--441.
\newblock \href {http://arxiv.org/abs/hep-ph/0404071}
  {\path{arXiv:hep-ph/0404071}}, \href
  {http://dx.doi.org/10.1016/j.physletb.2004.06.063}
  {\path{doi:10.1016/j.physletb.2004.06.063}}.

\bibitem{Actis:2008ug}
S.~Actis, G.~Passarino, C.~Sturm, S.~Uccirati, {NLO electroweak corrections to
  Higgs boson production at hadron colliders}, Phys.~Lett. B~670 (2008) 12--17.
\newblock \href {http://arxiv.org/abs/0809.1301} {\path{arXiv:0809.1301}},
  \href {http://dx.doi.org/10.1016/j.physletb.2008.10.018}
  {\path{doi:10.1016/j.physletb.2008.10.018}}.

\bibitem{Brein:2003wg}
O.~Brein, A.~Djouadi, R.~Harlander, {NNLO QCD corrections to the
  Higgs-strahlung processes at hadron colliders}, Phys.~Lett. B~579 (2004) 149.
\newblock \href {http://arxiv.org/abs/hep-ph/0307206}
  {\path{arXiv:hep-ph/0307206}}, \href
  {http://dx.doi.org/10.1016/j.physletb.2003.10.112}
  {\path{doi:10.1016/j.physletb.2003.10.112}}.

\bibitem{Ciccolini:2003jy}
M.~L. Ciccolini, S.~Dittmaier, M.~Kramer, {Electroweak radiative corrections to
  associated $WH$ and $ZH$ production at hadron colliders}, Phys. Rev. D~68
  (2003) 073003.
\newblock \href {http://arxiv.org/abs/hep-ph/0306234}
  {\path{arXiv:hep-ph/0306234}}, \href
  {http://dx.doi.org/10.1103/PhysRevD.68.073003}
  {\path{doi:10.1103/PhysRevD.68.073003}}.

\bibitem{Beenakker:2002nc}
W.~Beenakker, et~al., {NLO QCD corrections to $t \bar{t} H$ production in
  hadron collisions.}, Nucl. Phys. B~653 (2003) 151--203.
\newblock \href {http://arxiv.org/abs/hep-ph/0211352}
  {\path{arXiv:hep-ph/0211352}}, \href
  {http://dx.doi.org/10.1016/S0550-3213(03)00044-0}
  {\path{doi:10.1016/S0550-3213(03)00044-0}}.

\bibitem{Dawson:2003zu}
S.~Dawson, C.~Jackson, L.~H. Orr, L.~Reina, D.~Wackeroth, {Associated Higgs
  production with top quarks at the Large Hadron Collider: NLO QCD
  corrections}, Phys. Rev. D~68 (2003) 034022.
\newblock \href {http://arxiv.org/abs/hep-ph/0305087}
  {\path{arXiv:hep-ph/0305087}}, \href
  {http://dx.doi.org/10.1103/PhysRevD.68.034022}
  {\path{doi:10.1103/PhysRevD.68.034022}}.

\bibitem{Djouadi:1997yw}
A.~Djouadi, J.~Kalinowski, M.~Spira, {HDECAY: A program for Higgs boson decays
  in the standard model and its supersymmetric extension}, Comput. Phys.
  Commun. 108 (1998) 56--74.
\newblock \href {http://arxiv.org/abs/hep-ph/9704448}
  {\path{arXiv:hep-ph/9704448}}, \href
  {http://dx.doi.org/10.1016/S0010-4655(97)00123-9}
  {\path{doi:10.1016/S0010-4655(97)00123-9}}.

\bibitem{Bredenstein:2006rh}
A.~Bredenstein, A.~Denner, S.~Dittmaier, M.~M. Weber, {Precise predictions for
  the Higgs-boson decay $H\rightarrow WW/ZZ \rightarrow 4$leptons}, Phys. Rev.
  D~74 (2006) 013004.
\newblock \href {http://arxiv.org/abs/hep-ph/0604011}
  {\path{arXiv:hep-ph/0604011}}, \href
  {http://dx.doi.org/10.1103/PhysRevD.74.013004}
  {\path{doi:10.1103/PhysRevD.74.013004}}.

\bibitem{Actis:2008ts}
S.~Actis, G.~Passarino, C.~Sturm, S.~Uccirati, {NNLO computational techniques:
  the cases $H \to \gamma \gamma$ and $H \to g g$}, Nucl. Phys. B~811 (2009)
  182--273.
\newblock \href {http://arxiv.org/abs/0809.3667} {\path{arXiv:0809.3667}},
  \href {http://dx.doi.org/10.1016/j.nuclphysb.2008.11.024}
  {\path{doi:10.1016/j.nuclphysb.2008.11.024}}.

\bibitem{Dicus:2013lta}
D.~A. Dicus, C.~Kao, W.~W. Repko, {Comparison of $H\to\ell\bar{\ell}\gamma$ and
  $H\to\gamma\,Z, Z\to\ell\bar{\ell}$ including the ATLAS cuts. }\href
  {http://arxiv.org/abs/1310.4380} {\path{arXiv:1310.4380}}.

\bibitem{Chen:2012ju}
L.-B. Chen, C.-F. Qiao, R.-L. Zhu, {Reconstructing the 125 GeV SM Higgs boson
  through $\ell\bar{\ell}\gamma$. }\href {http://arxiv.org/abs/1211.6058}
  {\path{arXiv:1211.6058}}.

\bibitem{Firan:2007tp}
A.~Firan, R.~Stroynowski, {Internal conversions in Higgs decays to two
  photons}, Phys. Rev. D~76 (2007) 057301.
\newblock \href {http://arxiv.org/abs/0704.3987} {\path{arXiv:0704.3987}},
  \href {http://dx.doi.org/10.1103/PhysRevD.76.057301}
  {\path{doi:10.1103/PhysRevD.76.057301}}.

\bibitem{jimmy}
J.~M. Butterworth, J.~R. Forshaw, M.~H. Seymour, {Multiparton interactions in
  photoproduction at HERA}, Z. Phys. C~72 (1996) 637--646.
\newblock \href {http://arxiv.org/abs/hep-ph/9601371}
  {\path{arXiv:hep-ph/9601371}}, \href
  {http://dx.doi.org/10.1007/s002880050286} {\path{doi:10.1007/s002880050286}}.

\bibitem{ATLASsimulation}
{ATLAS Collaboration}, {The ATLAS simulation infrastructure}, Eur.~Phys.~J.~C
  70 (2010) 823--874.
\newblock \href {http://arxiv.org/abs/1005.4568} {\path{arXiv:1005.4568}},
  \href {http://dx.doi.org/10.1140/epjc/s10052-010-1429-9}
  {\path{doi:10.1140/epjc/s10052-010-1429-9}}.

\bibitem{geant}
S.~Agostinelli, et~al., Geant4 - a simulation toolkit, Nucl.~Instrum.~Methods~A
  506 (2003) 250.
\newblock \href {http://dx.doi.org/10.1016/S0168-9002(03)01368-8}
  {\path{doi:10.1016/S0168-9002(03)01368-8}}.

\bibitem{muonreco}
{ATLAS Collaboration}, {Muon reconstruction efficiency in reprocessed 2010 LHC
  proton-proton collision data recorded with the ATLAS detector},
  {ATLAS-CONF-2011-063} (2010) \url{http://cds.cern.ch/record/1345743}.

\bibitem{PhotonPerformanceNote}
{ATLAS Collaboration}, {Expected photon performance in the ATLAS experiment},
  {ATLAS-PHYS-PUB-2011-007 (2011),}  \url{http://cds.cern.ch/record/1345329}.

\bibitem{GSFConf}
{ATLAS Collaboration}, {Improved electron reconstruction in ATLAS using the
  Gaussian Sum Filter-based model for bremsstrahlung}, {ATLAS-CONF-2012-047
  (2012),}  \url{http://cds.cern.ch/record/1449796}.

\bibitem{Aad:2011mk}
{ATLAS Collaboration}, {Electron performance measurements with the ATLAS
  detector using the 2010 LHC proton-proton collision data}, Eur.~Phys.~J. C 72
  (2012) 1909.
\newblock \href {http://arxiv.org/abs/1110.3174} {\path{arXiv:1110.3174}},
  \href {http://dx.doi.org/10.1140/epjc/s10052-012-1909-1}
  {\path{doi:10.1140/epjc/s10052-012-1909-1}}.

\bibitem{Aad:2010sp}
{ATLAS Collaboration}, {Measurement of the inclusive isolated photon cross
  section in $pp$ collisions at $\sqrt{s}= 7\TeV$ with the ATLAS detector},
  Phys.~Rev.~D 83 (2011) 052005.
\newblock \href {http://arxiv.org/abs/1012.4389} {\path{arXiv:1012.4389}},
  \href {http://dx.doi.org/10.1103/PhysRevD.83.052005}
  {\path{doi:10.1103/PhysRevD.83.052005}}.

\bibitem{Aad:2013izg}
{ATLAS Collaboration}, {Measurements of $W\gamma$ and $Z\gamma$ production in
  $pp$ collisions at $\sqrt{s}$= 7 TeV with the ATLAS detector at the LHC},
  Phys.~Rev. D 87 (2013) 112003.
\newblock \href {http://arxiv.org/abs/1302.1283} {\path{arXiv:1302.1283}},
  \href {http://dx.doi.org/10.1103/PhysRevD.87.112003}
  {\path{doi:10.1103/PhysRevD.87.112003}}.

\bibitem{ATLAS-CONF-2012-123}
{ATLAS Collaboration}, \href{http://cds.cern.ch/record/1473426/}{{Measurements
  of the photon identification efficiency with the ATLAS detector using 4.9
  fb$^{-1}$ of $pp$ collision data collected in 2011}}, ATLAS-CONF-2012-123.
\newline\urlprefix\url{http://cds.cern.ch/record/1473426/}

\bibitem{HiggsCombStatMethod2011}
{ATLAS and CMS Collaborations}, {Procedure for the LHC Higgs boson search
  combination in Summer 2011}, ATLAS-PHYS-PUB-2011-011, CMS NOTE-2011/005.

\bibitem{cls}
A.~L. Read, {Presentation of search results: The CL(s) technique}, J.~Phys.~G:
  Nucl. Part. Phys 28 (2002) 2693.
\newblock \href {http://dx.doi.org/10.1088/0954-3899/28/10/313}
  {\path{doi:10.1088/0954-3899/28/10/313}}.

\bibitem{Cowan:2010js}
G.~Cowan, K.~Cranmer, E.~Gross, O.~Vitells, {Asymptotic formulae for
  likelihood-based tests of new physics}, Eur.~Phys.~J. C~71 (2011) 1554.
\newblock \href {http://arxiv.org/abs/1007.1727} {\path{arXiv:1007.1727}},
  \href {http://dx.doi.org/10.1140/epjc/s10052-011-1554-0}
  {\path{doi:10.1140/epjc/s10052-011-1554-0}}.

\end{thebibliography}

\onecolumn
\clearpage

\begin{flushleft}
{\Large The ATLAS Collaboration}

\bigskip

G.~Aad$^{\rm 84}$,
T.~Abajyan$^{\rm 21}$,
B.~Abbott$^{\rm 112}$,
J.~Abdallah$^{\rm 152}$,
S.~Abdel~Khalek$^{\rm 116}$,
O.~Abdinov$^{\rm 11}$,
R.~Aben$^{\rm 106}$,
B.~Abi$^{\rm 113}$,
M.~Abolins$^{\rm 89}$,
O.S.~AbouZeid$^{\rm 159}$
% ...
\bigskip
\\
% ...
$^{11}$ Institute of Physics, Azerbaijan Academy of Sciences, Baku, Azerbaijan\\
% ...
$^{21}$ Physikalisches Institut, University of Bonn, Bonn, Germany\\
% ...
$^{84}$ CPPM, Aix-Marseille Universit{\'e} and CNRS/IN2P3, Marseille, France\\
% ...
$^{89}$ Department of Physics and Astronomy, Michigan State University, East Lansing MI, United States of America\\
% ...
$^{106}$ Nikhef National Institute for Subatomic Physics and University of Amsterdam, Amsterdam, Netherlands\\
% ...
$^{112}$ Homer L. Dodge Department of Physics and Astronomy, University of Oklahoma, Norman OK, United States of America\\
$^{113}$ Department of Physics, Oklahoma State University, Stillwater OK, United States of America\\
% ...
$^{116}$ LAL, Universit{\'e} Paris-Sud and CNRS/IN2P3, Orsay, France\\
% ...
$^{152}$ Institute of Physics, Academia Sinica, Taipei, Taiwan\\
% ...
$^{159}$ Department of Physics, University of Toronto, Toronto ON, Canada\\
% ...
\end{flushleft}


\end{document}